\documentclass[amssymb,prb,twocolumn,showpacs,superscriptaddress]{revtex4}
\usepackage{amsmath}
\usepackage{graphicx}
\usepackage{verbatim}
\usepackage{graphics}
\usepackage{units}
\usepackage{color}
\usepackage{ulem}
\usepackage[colorlinks=true,citecolor=blue,urlcolor=blue,linkcolor=black]{hyperref}
\usepackage{bm}
\usepackage{slashed}
\usepackage{braket} 
\usepackage{longtable}
\usepackage{xcolor}
\usepackage{amssymb}
\usepackage{appendix}
\usepackage{epstopdf}
\usepackage{tikz}
\usetikzlibrary{decorations.pathmorphing}
\usetikzlibrary{decorations.markings}
\usepackage{overpic}
\usetikzlibrary{patterns}
\usetikzlibrary{shapes,snakes}
\usetikzlibrary{arrows}

\newcommand{\eq}[1]{Eq. \eqref{#1}}

\begin{document}
\title{
Coupled Landau-Zener-St\"uckelberg Quantum Dot Interferometers
}
\author{Fernando Gallego-Marcos}
\affiliation{Instituto de Ciencia de Materiales de Madrid, CSIC, Cantoblanco, 28049, Madrid, Spain}
\affiliation{Institut f\"ur Festk\"orperphysik, Leibniz Universit\"at Hannover, Appelstrasse 2, 30167 Hannover, Germany}
\author{Rafael S\'anchez}
\affiliation{Instituto de Ciencia de Materiales de Madrid, CSIC, Cantoblanco, 28049, Madrid, Spain}
\affiliation{Instituto Gregorio Mill\'an, Universidad Carlos III de Madrid, 28911 Legan\'es, Madrid, Spain}
\author{Gloria Platero}
\affiliation{Instituto de Ciencia de Materiales de Madrid, CSIC, Cantoblanco, 28049, Madrid, Spain}
\begin{abstract}
We investigate the interplay between long range and direct photo-assisted transport in a triple quantum dot chain where local ac voltages are applied to the outer dots.  
We propose the phase difference between the two ac voltages as external parameter, which can be easily tuned to manipulate the current characteristics. For gate voltages in phase opposition
we find 
quantum destructive interferences 
analogous to the interferences in closed loop undriven triple dots. As the voltages oscillate in phase,  interferences between multiple 
paths give rise to dark states.
Those totally cancel the current, and could be experimentally resolved. 
\end{abstract}
\maketitle

%
%
\section{Introduction}\label{Sec::Introduction}
A system that is driven non-adiabatically through the avoided crossing of two states undergoes a 
transition~\cite{landau,zener,stuckelberg,majorana}. The probability of the transition depends on the parameters of the driving and the splitting at the crossing. The latest is given by the coupling between the diabatic states. Repeating the passing through the crossing introduces different paths to end in a given state, which gives rise to constructive interference.
The control of this mechanism in solid state qubits has become a standard tool in the manipulation of quantum states~\cite{oliver,petta,gaudreau}, the generation of entanglement~\cite{quintana}, or the measurement of the qubit coherence timescales~\cite{foster}. 

In periodically driven quantum dot systems, this effect is measured as photon-assisted tunneling resonances~\cite{gloria}. An electron is hence delocalized between tunnel coupled quantum dots when the detuning of their energy levels is a multiple of the driving frequency $n\hbar\omega$~\cite{oosterkamp}. The tunnel coupling is renormalized by the ac field by a factor which depends on the amplitude 
and frequency of the driving~\cite{shevchenko,hanggi}. 
Recently, striking electron spin resonance measurements in quantum dot systems~\cite{stehlik} have been interpreted in terms of multilevel crossings~\cite{danon}. 
Three-level crossings may also
lead to peculiar phenomena such as dark resonances~\cite{renzoni,darkbell}. 

Triple quantum dots (TQDs) are ideal systems  for investigation of such processes. 
On one hand, the spacial separation of three states~\cite{rogge,ghislain}, one in each dot (L, C and R) makes it possible to manipulate them individually by means of gate voltages~\cite{braakman}. Hence, different drivings can be applied to the different levels by applying localized time-dependent gate voltages to each quantum dot~\cite{kiselev}. Thus, not only the amplitude and frequency of the driving~\cite{grifoni,gloria}, but also phase differences~\cite{alvaro} become important. 

On the other hand, the tunnel coupling between all three states can be tuned, also between those that are not directly coupled. 
Indeed, long-range transport betweeen the edge dots of a linear TQD has been very recently detected~\cite{busl,floris,spin}. During these higher-order (cotunneling) transitions, the center dot is only virtually occupied. Hence they involve the direct transfer of a charge or a spin qubit betweeen distant sites, avoiding decoherence and relaxation in the intermediate region. 



\begin{figure}[b]
\includegraphics[width=0.9\linewidth] {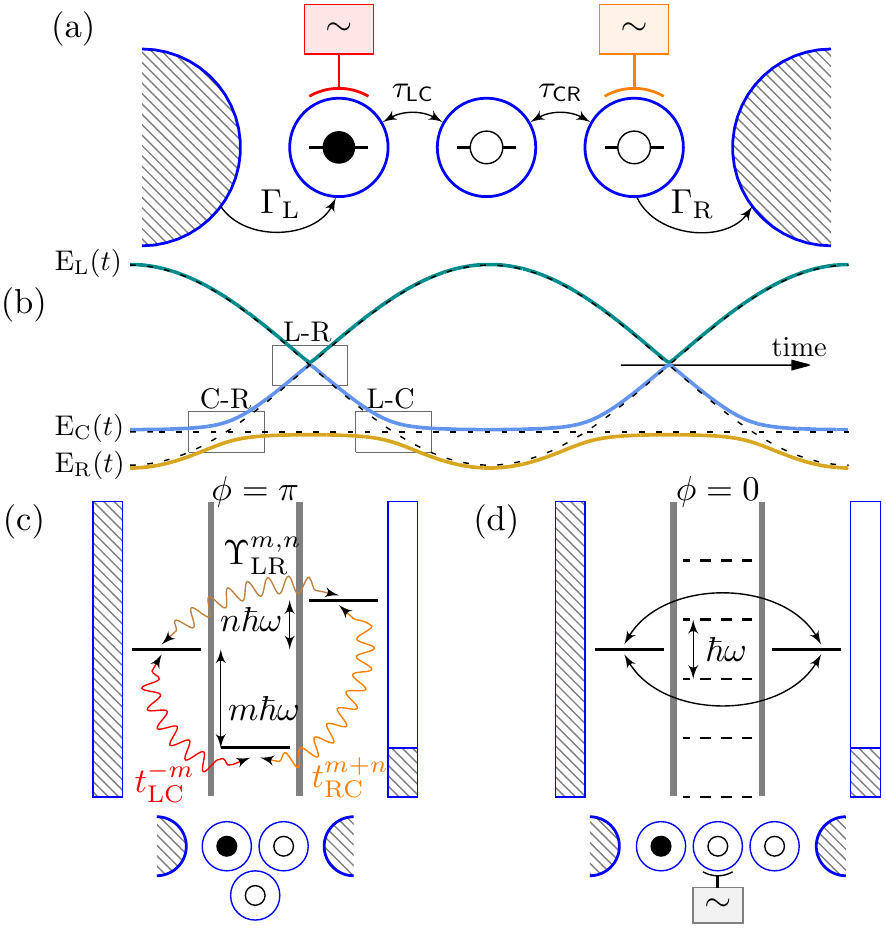}
\caption{ (a) TQD in series connected to leads. Two ac voltages are applied to the outer dots, with a phase difference $\phi$. (b) Time evolution of the energy levels for the left, center and right dots for $\phi=\pi$, showing the different crossings mediated by direct (L-C and C-R) and virtual tunneling (L-R).  (c) For $ \phi = \pi $ the driving induces resonant transitions (either direct, $t_{ik}^n$ or virtual, $\Upsilon_\text{LR}^{m,n}$) between all the levels. 
The system can thus be mapped to an undriven triangular TQD depicted below.
(d) The case where the edge dot levels oscillate in phase ($ \phi=0$) is equivalent to that where only the center dot is driven. Transport is then governed by sidebands in the center dot. 
 }
\label{Fig:esquema}
\end{figure}


By applying sinusoidal signals to the outer dot gates, the system can be driven through anticrossings among the three states, performing Landau-Zener transitions: L-C, C-R and L-R, cf. Fig.~\ref{Fig:esquema}(b). Note that the long-range L-R transition is paralell to the direct-tunneling L-C, C-R trajectory. The system then behaves as a combination of coupled interferometers. The interference patterns, coming from real or virtual paths, can be detected by coupling the system to source and drain contacts and measuring the current that passes through it along a voltage bias $eV$, as sketched in Fig.~\ref{Fig:esquema}.
The versatility of this system makes it suitable for investigating many different configurations. Here we will consider the case where the two drivings are equal in amplitude and frequency, $ V_{l}(t)=(V_{\rm ac}/2)\cos(\omega t+\phi_l)$, but they have a phase difference $\phi=\phi_{\rm R}-\phi_{\rm L}$. 
As we show below, the dependence on the phase is of paramount importance and it has not yet been addressed. 

We concentrate on two cases, when the dots oscillate with opposite ($\phi=\pi$) or the same phase ($\phi=0$).
In the first case,  $\phi=\pi$, crossings of all three states occur. Direct tunneling transitions between left and center and center and right  take place when the corresponding energy levels satisfy $\varepsilon_{\rm C}-\varepsilon_{\rm L}\approx m\hbar\omega$ and $\varepsilon_{\rm R}-\varepsilon_{\rm C}\approx m'\hbar\omega$. Importantly, even if they are not directly coupled, the crossing of the outer dot levels induces a long-range tunneling transition. It is mediated by the virtual occupation of the center dot, when $\varepsilon_{\rm R}-\varepsilon_{\rm L}\approx n\hbar\omega$, regardless of the energy $\varepsilon_{\rm C}$~\cite{aguado,flensberg,fernando,stano}. Thus, different paths are possible for left to right transport [see Fig.~\ref{Fig:esquema}(c)] remarkably leading to the interference of real and virtual (cotunneling) transitions.

Differently if the two dots oscillate in phase ($\phi=0$), only L-C and C-R crossings are possible. The system is therefore equivalent to the one where only the center dot is driven. Hence $\varepsilon_{\rm C}$ develops sidebands separated by an energy $\hbar\omega$, as shown in Fig.~\ref{Fig:esquema}(d). Resonant transitions involving the sidebands is the main transport mechanism. However, long-range tunneling is also possible if the outer levels have the same energy.
The virtual occupation of sidebands with either positive or negative detuning causes unexpected interferences, similar to two path interferometers~\cite{alfredo,superexchange}.

The paper is organized as follows: we present the model and numerical results in Sec.~\ref{sec::model}. An analytical approach is performed in Sec.~\ref{sec::anal} which allows us to give an interpretation of the main features in Sec.~\ref{sec::inter}. A summary is given in Sec.~\ref{sec::conc}.


\section{Model}\label{sec::model}
We assume the system to be in the Coulomb blockade regime, where only one electron is allowed in the TQD at a time. In such a case, the spin of the electron does not play any role and we can ignore it. The states of our Hamiltonian are defined in the basis \{$\ket{L}$, $\ket{C}$, $\ket{R}$, $\ket{0}$\}, where $\ket{i}$ represents one electron in the \textit{i-dot}, and $\ket{0}$ corresponds to the case where the system is empty. Furthermore, we consider a large bias voltage such that electrons flow unidirectionally from the left to the right reservoir.
We write the Hamiltonian of the system $\hat{H}(t)=\hat{H}_0+\hat{H}_{\text{lead}}+\hat{H}_{\text{int}}+\hat{H}_{\text{ac}}(t)$, where
\begin{align}
\hat{H}_{0}=\hat{H}_\varepsilon{+}\hat{H}_\tau=\sum_{i}\varepsilon_i\hat{c}^\dag_i\hat{c}_i {+}\sum_i(\tau_{i,i+1}\hat{c}^\dag_i\hat{c}_{i+1}+\text{H.c.})
\end{align}
describes the TQD with energy levels $\varepsilon_i$ localized in each dot and (nearest neighbor) interdot tunnel couplings $\tau_{ij}$. The term $\hat{H}_{\text{lead}}=\sum_{lk}\varepsilon_{lk}\hat{d}^\dag_{lk}\hat{d}_{lk}$ represents the leads, and $\hat{H}_{\text{int}}=\sum_{lki}\gamma_{l}\hat{d}^\dag_{lki}\hat{c}_i+\text{H.c.}$ is the system-lead interaction Hamiltonian, given by the coupling  $\gamma_{l}$. 
The driving $\hat{H}_{\text{ac}}(t)=\sum_lV_l(t)\hat{c}^\dag_l\hat{c}_l$ acts on the edge dots $l$=L,R.

In order to treat the explicit time dependence, it is convenient to perform a unitary transformation $\hat{H}'(t)=\hat{U}^\dag(t)\left[\hat{H}(t)-i\hbar\partial_t\right]\hat{U}(t)$, where
\begin{align}\hat{U}(t)=\exp\left[\frac{i}{\hbar}\int_0^t\text{d}t_1\hat{H}_{\text{ac}}(t_1)\right].
\end{align}
It leaves the diagonal terms time independent. The system Hamiltonian transforms as $\hat{H}'_0(t)=\hat{H}_\epsilon+\hat{H}'_{\tau}$, with
\begin{align}
\hat{H}'_{\tau}&=\sum_{\nu=-\infty}^\infty t_{LC}^\nu \text{e}^{i\nu\omega t}\hat{c}^\dag_C\hat{c}_L+t_{CR}^\nu \text{e}^{-i\nu\omega t}\hat{c}^\dag_R\hat{c}_C+\text{H.c.}\label{eq::tauTrans}
\end{align}
where the hopping terms $t_{LC}^\nu=\tau_{LC}J_\nu(V_{\text{ac}}/2\hbar\omega)$, $t_{CR}^\nu(\phi)=\tau_{CR}J_\nu(V_{\text{ac}}/2\hbar\omega)e^{-i\nu\phi}$ are renormalized by the $\nu$-th order Bessel functions $J_{\nu}(x)$. In what follows, we will simply write $J_\nu$, unless its argument is different from $V_{\text{ac}}/(2\hbar\omega)$.
In the large bias regime we are considering, $eV\gg\hbar\omega,V_\text{ac}$, the effect of the transformation on the tunneling term $\hat{H}_{\text{int}}$ can be disregarded due to the normalization of Bessel functions.

In the weak coupling limit and assuming the Born-Markov secular approximation~\cite{OpenQuantumSys},
the quantum master equation for the reduced density matrix of the TQD becomes~\cite{stoof}:
\begin{align}
\dot{\rho}(t)=\frac{i}{\hbar}[\hat{H}_\varepsilon+\hat{H}'_\tau(t),\rho(t)]+(\mathcal{L}_\Gamma-\mathcal{L}_{\Lambda})\rho(t)\label{eq::Master}
\end{align}
where 
\begin{align}
\braket{m|\mathcal{L}_{\Gamma}\rho(t)|n}&=\sum_{k\neq n}(\Gamma_{nk}-\Gamma_{kn})\rho_{mn}(t)\delta_{mn},\\
\braket{m|\mathcal{L}_\Lambda\rho(t)|n}&=\frac{1}{2}\left(\sum_{k\ne n}\Gamma_{kn}{+}\sum_{k\neq m}\Gamma_{km}\right)\rho_{mn}(1{-}\delta_{mn}).\nonumber
\end{align}
In the large bias regime, only two rates participate, $\Gamma_\text{L0}=\Gamma_\text{L}$ and $\Gamma_\text{0R}=\Gamma_\text{R}$, with $\Gamma_l=(2\pi/\hbar){\cal D}_l|\gamma_l|^2$, and ${\cal D}_l$ being the density of states in the lead to which dot $l$ is coupled.
The steady state occupations $\rho^{\text{st}}_{ii}$ are obtained by numerically integrating in time Eq.~\eqref{eq::Master}. With those, we get the stationary current, given by $I=e\Gamma_\text{R}\rho_\text{RR}^{\text{st}}$.

\begin{figure}[t]
\includegraphics[width=1.0\linewidth] {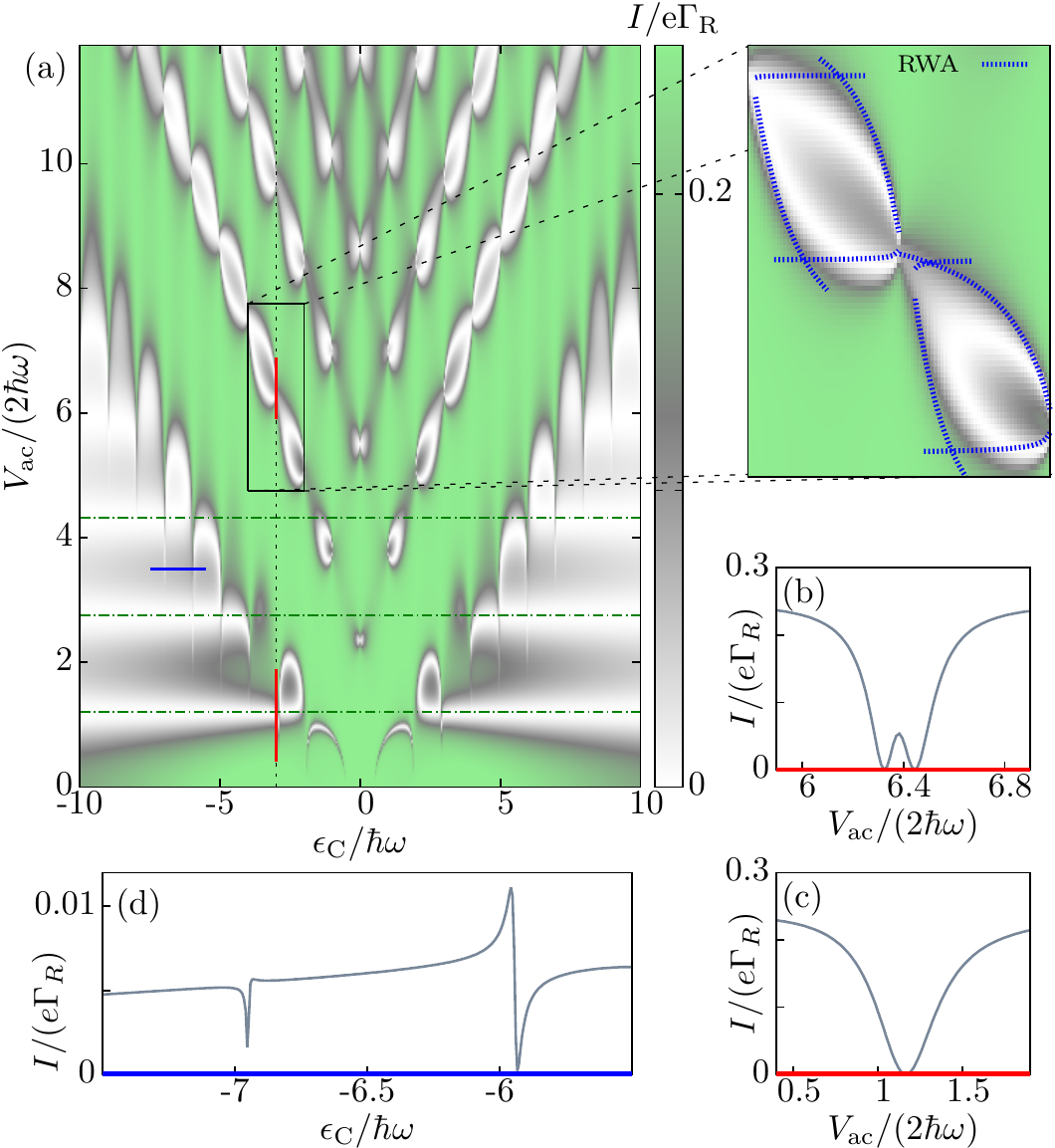}
\caption{(a) Current through a TQD by tuning $\varepsilon_\text{C}$ and  $V_{\text{ac}}$. $\epsilon_{\text{R}}=\epsilon_{\text{L}}=0$, $\omega= 3.3\tau_\text{LC}$, $\tau_\text{LC}=\tau_\text{CR}$  and $\phi=\pi$. Superimposed to the well-known Landau-Zener-St\"uckelberg pattern, we find exact cancellations of the current that are attributed to the contribution of virtual tunneling transitions. The green dashed lines correspond to the zeros of  $J_0(V_{\text{ac}}/\hbar\omega)$, cf. Eq. \eqref{eq::RWA2H_simp}.
(b)  and $($c)  show the  current for two cuts along the resonance $\varepsilon_\text{C}=-3\hbar\omega$ [marked as red lines in (a)]. While in $($c) the current vanishes coinciding with $J_0(V_{\rm ac}/\hbar\omega)=0$, a double suppression appears in $($b) around the first zero of $J_3(V_{\rm ac}/2\hbar\omega)$. The complicated double-minimum  interference patterns are of the same nature as the ones observed in an undriven triangular configuration. By means of the RWA one can derive an analytical condition Eq. \eqref{eq::condDS} [blue dashed lines in the inset of (a)] for the dark state Eq. \eqref{eq::DSPI}. (d) current for $V_\text{ac}/(2\hbar\omega)=3.51$ in the cotunnel regime,  where Fano-like resonances are observed.}
\label{fig::transportLZS}
\end{figure}

\subsection{Numerical observations}\label{sec::NumObs}
\begin{figure}[t]
\includegraphics[width=1.0\linewidth] {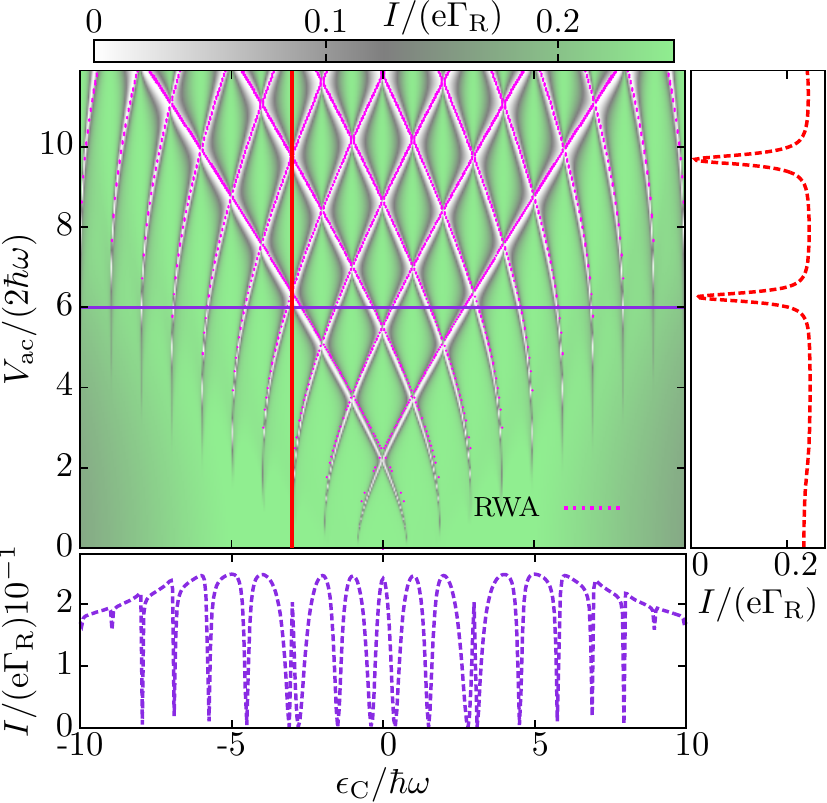}
\caption{
Current through the TQD as a function of the detuning $\varepsilon_\text{C}$ and $V_{\text{ac}}$ for $\phi=0$, with $\epsilon_{\text{R}}=\epsilon_{\text{L}}=0$, $\omega= 3.3\tau_\text{LC}$, $\tau_\text{LC}=\tau_\text{CR}$. 
Narrow antiresonances are due to destructive interference between sidebands with positive and negative detuning with respect to the L-R resonance. 
The dashed lines  within the white regions correspond to the fulfillment of condition \eqref{eq:DSAC} for dark states, see text below. Right panel: current at fixed $\varepsilon_\text{C}=-3\hbar\omega$ (red vertical line). The dips correspond to the zeros of $J_3$. Bottom panel:  current cut along $V_\text{ac}=3\hbar\omega$ (horizontal blue line) showing a symmetric configuration of zeros which comes from the interference between sidebands.}
\label{fig::comparacion}
\end{figure}

The current for the different configurations, $\phi=\pi$ and $\phi=0$, is plotted in Figs.~\ref{fig::transportLZS} and \ref{fig::comparacion} respectively, as functions of the driving amplitude, $V_{\rm ac}$, and the detuning of the centre dot level $\varepsilon_{\rm C}$ from $\varepsilon_{\rm L}=\varepsilon_{\rm R}=0$. 
In both configurations, we can clearly distinguish two regions by the ratio $\alpha=V_{\rm ac}/(2|\varepsilon_{\rm C}|)$, which marks the onset of photon-assisted transitions. The current shows very different features in each region, which we detail below. 

For $\phi=\pi$, the current is cancelled for certain driving amplitudes in the regime $\alpha<1$ where the center dot is not excited, see horizontal lines in Fig.~\ref{fig::transportLZS}. For larger amplitudes, we observe the expected photon-assisted resonances around $\varepsilon_{\rm C}=m\hbar\omega$ which are suppressed at the $i$-th zeros $z_{i,m}$ of $J_m$~\cite{shevchenko}. More surprisingly, we find additional off-resonance pairs of suppressions of the current in between zeros of consecutive resonances, cf. inset in Fig.~\ref{fig::transportLZS}. Remarkably, the cancellation of the current is exact.
These sharp features are a manifestation of the interference of real and virtual transitions, as we discuss in the next section.

Very differently, for $\phi=0$, no significant features are observed below $\alpha=1$, cf. Fig.~\ref{fig::comparacion}. For larger amplitudes, we find narrow  off-resonance current suppressions all the way between the $i$-th zeros $z_{i,m}$ of increasing/decreasing order, e.g. current is cancelled between $z_{i,m}$ and $z_{i,m\pm1}$, cf. Fig.~\ref{fig::comparacion}.
 These features come from the destructive interference between sidebands with positive and negative detuning with respect to the L-R resonance.

\section{Analytical approach}\label{sec::anal}
We perform an analytical treatment consisting on a perturbative expansion of the evolution operator in the inter-dot tunneling, and a rotating wave approximation (RWA) close to the relevant resonances. Direct tunneling between nearby dots (L-C and C-R) is contained in the first order of the expansion. The long-range tunneling between the left and right dots is only captured in the second order. 
We are interested in the configurations where both real and virtual transitions coexist. This occurs close to the double resonance conditions $\varepsilon_C-\varepsilon_L\approx m\hbar\omega$ and $\varepsilon_R-\varepsilon_L\approx n\hbar\omega$, as depicted in Fig.~\ref{Fig:esquema}. 
Close to these, one can transform the Hamiltonian to the rotating frame and  neglect the fast oscillating terms of the Hamiltonian. With this RWA we obtain a time independent Hamiltonian:
\begin{align}
\hat{H}_{\text{RWA}}^{m,n}(\phi){=}\left(\begin{array}{ccc}\tilde{\varepsilon}_{\text{L}}^{n}{+}\Lambda_{\text{LL}}^{m,n}&t_{\text{LC}}^{-m}& \Upsilon_{\text{LR}}^{m,n^*}(\phi)
\\\ &\ &\ \\ t_{\text{CL}}^{-m}&\tilde{\varepsilon}_{\text{C}}^{m+n}{+}\Lambda_{\text{CC}}^{m,n}&t_{\text{CR}}^{-m-n}(\phi)\\\ &\ &\ \\ \Upsilon_{\text{LR}}^{m,n}(\phi)&t_{\text{RC}}^{-m-n}(\phi)&\tilde{\varepsilon}_{\text{R}}^{0}{+}\Lambda_{\text{RR}}^{m,n}\\\end{array}\right)\label{eq::RWA2H}
\end{align}
where	 $\tilde{\varepsilon}_{i}^{n}=\varepsilon_i+n\hbar\omega$. $\Lambda_{\text{LL}}^{m,n}=\tau_{\rm LC}^2\lambda_m$, $\Lambda_{\text{RR}}^{m,n}=\tau_{\rm CR}^2\lambda_{m+n}$ and $\Lambda_{\text{CC}}^{m,n}=-(\Lambda_{\text{LL}}^{m,n}+\Lambda_{\text{RR}}^{m,n})$ are the level shifts coming from the second order tunnel contribution, with $\lambda_p=\sum_{\nu}(J_\nu-J_p)J_\nu/[(p-\nu)\hbar\omega]$.  More importantly, 
\begin{align}
\Upsilon_{\text{LR}}^{m,n}(\phi)&=\tau_{\text{LC}}\tau_{\text{CR}}{\sum_{\nu=-\infty}^\infty}J_{\nu}\left[\frac{(J_{\nu+n}-J_{m+2n})e^{i\nu\phi}}{2(\nu-m-n)\hbar\omega}\right.\nonumber\\
&\left.+\frac{\left( J_{\nu-n}e^{i\nu\phi} -J_{m-n}e^{im\phi}\right) e^{-in\phi}}{2(\nu-m)\hbar\omega}\right]\label{eq::para3}
\end{align}
describes the second-order photon-assisted tunneling between the outermost dots. 
%
%
%
%

\section{Interpretation}\label{sec::inter}
We emphasize the role of the phase difference in the off-diagonal elements in Eq. \eqref{eq::RWA2H}. As we discuss in the following, the origin of the destructive interference patterns is very different in the two configurations of interest here, $\phi=0,\pi$.

\subsection{Opposite phase drivings, $\phi=\pi$}
\label{sec::phipi}
Let us start by the case $\phi=\pi$, plotted in Fig.~\ref{fig::transportLZS}.
The double dip detailed in the inset can be understood directly by looking at the structure of the effective Hamiltonian in \eq{eq::RWA2H}. The long-range coupling of left and right dots makes it analogous with a TQD in a closed-loop triangular configuration. Such a system is known to lead to dark states when different paths around the triangle interfere destructively~\cite{michaelis}. This occurs when particular relations between couplings and detuning are met~\cite{clive}. In our case, with $\tilde{\epsilon}_{\text{L},0}=\tilde{\epsilon}_{\text{R},0}=\tilde{\epsilon}_{\text{C},m}$, we find the following condition for a dark state:
\begin{align}
\Delta^{m,0}\Upsilon_{\text{LR}}^{m,0}(\phi)t_{\text{CR}}^{-m}(\phi)=t_{\text{LC}}^{-m}\big(\Upsilon_{\text{LR}}^{m,0}(\phi)^2-t_{\text{CR}}^{-m}(\phi)^2\big)
\label{eq::condDS}
\end{align}
where:
$\Delta^{m,0}=\Lambda_{\text{LL}}^{m,0}-\Lambda_{\text{CC}}^{m,0}$.
Eq. \eqref{eq::condDS} is the exact condition for the existence of a dark-state: 
\begin{align} 
 \ket{\Psi_{\text{DS}}}=N\left(t_{\text{CR}}^{-m}(\phi)\ket{L}-\Upsilon_{\text{LR}}^{m,0}(\phi)\ket{C}\right), \label{eq::DSPI}
\end{align} 
where N is a normalization constant.
Note that it contains no contribution of the state $\ket{\rm R}$, which is coupled to the drain lead. Therefore, no current flows through the system.  

The dark state condition~\eqref{eq::condDS}  is fulfilled for all the dips observed in Fig. \ref{fig::transportLZS}, showing that the linear  driven system behaves as an undriven triangular one around every $z_{i,m}$. The condition \eqref{eq::condDS} is represented by dashed lines in the inset of Fig.~\ref{fig::transportLZS} with a remarkable agreement. We can thus interpret these dips as the destructive interference between first- and higher-order tunneling transitions. 

When the ac amplitude is small, $\alpha<1$, we have $J_{m}\ll 1$, for $m>1$. Thus, in this regime the effective direct couplings $t_{\text{CL}}^{m},t_{\text{CR}}^{m+n}$ are negligible. This is not the case for the long-range coupling. Using the addition theorem for Bessel functions, it can be approximated to  $\Upsilon_{\text{LR}}^{m,n}\rightarrow-\tau_\text{LC}\tau_\text{CR}J_n{\left(\frac{V_{\text{ac}}}{\hbar\omega}\right)}/m\hbar\omega$. Note that the argument of the Bessel function is doubled and therefore it is not necessarily negligible. Hence, transport is dominated by cotunneling and Eq.~\eqref{eq::RWA2H} can be mapped to an effective two level system~\cite{fernando}:
\begin{equation}
\hat{H}_{\text{RWA}}^{m,n}(\pi)\approx(\tilde{\epsilon}_{\text{L},n}-\tilde{\epsilon}_{\text{R},0}){\boldsymbol\sigma}_z-J_n{\left(\frac{V_{\text{ac}}}{\hbar\omega}\right)}\frac{\tau_\text{LC}\tau_\text{CR}}{m\hbar\omega}{\boldsymbol\sigma}_x,
\label{eq::RWA2H_simp}
\end{equation}
with the Pauli matrices ${\boldsymbol\sigma}_i$. With this model, we can interpret the different cancellations of the current marked by horizontal dashed lines in Fig.~\ref{fig::transportLZS}. They are due to the zeros of $J_0{\left(\frac{V_{\text{ac}}}{\hbar\omega}\right)}$, therefore representing a long-range analog of coherent destruction of tunneling~\cite{grifoni}.

In this regime ($\alpha<1$), the center dot remains uncoupled from the background cotunneling transport. By increasing $V_\text{ac}$, the resonant photon-assisted L-C and C-R  transitions are activated. The center dot can then be considered as a localized state coupled to a continuum leading to asymmetric Fano-like resonances~\cite{kobayashi,entin}, see Fig.~\ref{fig::transportLZS}(d).  


\subsection{In-phase drivings, $\phi=0$} 
\label{sec::phi0}
We now consider the configuration with $\phi=0$. In this case Eq.~\eqref{eq::condDS} is only fulfilled for the trivial solution when all the couplings are zero.  As the driving does not induce the left and right levels to cross, cotunneling is only effective when $\varepsilon_\text{L}=\varepsilon_\text{R}$, i.e., $n=0$. In that case, the two levels cross the center one simultaneously. This situation can be mapped into having the driving in the center dot and sidebands at energies $\varepsilon_\text{C}+\nu\hbar\omega$, cf. Fig.~\ref{Fig:esquema}(d). If the center dot is detuned $\varepsilon_\text{C}=m\hbar\omega+\Delta$, the different sidebands will act as parallel channels for the cotunneling transition. The effective Hamiltonian then reads:
\begin{equation}
\hat{H}_{\text{RWA}}^{m,n}(0)\approx\Delta{\boldsymbol\sigma}_z+\tau_\text{LC}\tau_\text{CR}\sum_{\nu=-\infty}^\infty\Omega_{\nu,m}(\Delta){\boldsymbol\sigma}_x.
\label{eq::RWA2H0_simp}
\end{equation}
with the sideband-dependent couplings $\Omega_{\nu,m}(\Delta)=J_\nu^2/[(\nu-m)\hbar\omega-\Delta]$. Note that they can be tuned by the driving amplitude and the detuning.
If the condition 
\begin{align}
\sum_{\nu=-\infty}^m\Omega_{\nu,m}(\Delta)=-\sum_{\nu=m+1}^\infty\Omega_{\nu,m}(\Delta)
\label{eq:DSAC}
\end{align}
is met, sidebands with positive ($\nu>m$) and negative ($\nu<m$) detuning will destructively interfere. This leads to the narrow cancellations of the current shown in Fig.~\ref{fig::comparacion}. The condition \eqref{eq:DSAC} is represented there by dashed lines. We note that the agreement with the full numerics is excellent in all the configuration space. The system then behaves as an ac driven interferometer of multiple transitions.

\subsection{Arbitrary phase difference}
\begin{figure}[t]
\includegraphics[width=1.0\linewidth] {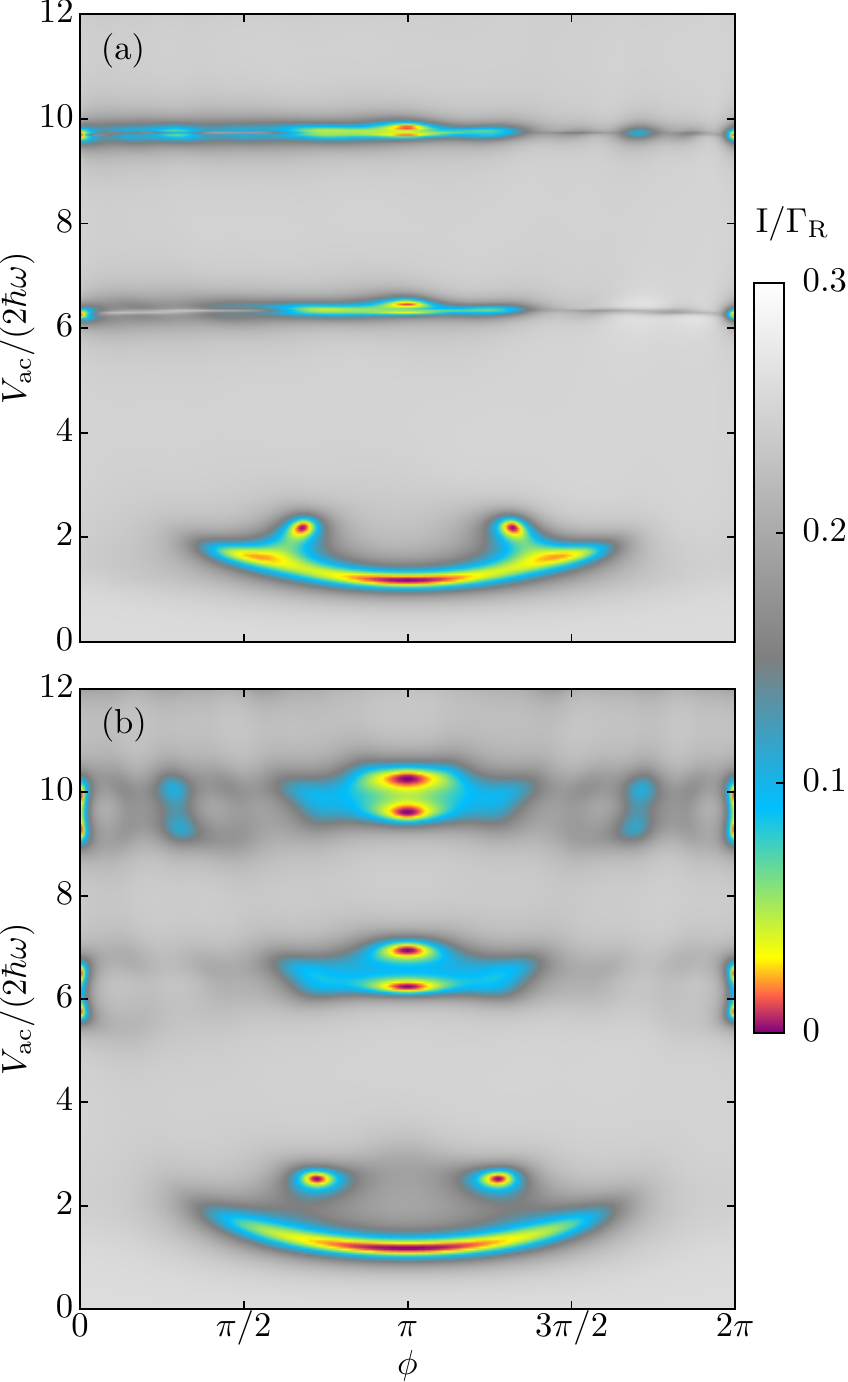}
\caption{
Current through the TQD numerically calculated with Eq. \eqref{eq::Master} by tuning $\phi$ and $\text{V}_\text{ac}$. In both plots  $\epsilon_{\text{R}}=\epsilon_{\text{L}}=0$, $\omega= 3.3\tau_\text{LC}$ and $\tau_\text{LC}=\tau_\text{CR}$. (a) The central dot is in resonance with the third side-band of the two outer dots, $\epsilon_{\text{C}}=\epsilon_{\text{L}}-3\hbar\omega$. A cut at $\phi=\pi$ corresponds to Fig. \ref{fig::transportLZS}(b) and (c), and at $\phi=0$ to the right vertical cut of Fig. \ref{fig::comparacion}. (b) Same as in (a) but with the central dot out of resonance with the outer dots, $\epsilon_{\text{C}}=\epsilon_{\text{L}}-3.2\hbar\omega$.}
\label{fig::fase}
\end{figure}

The effective Hamiltonian \eqref{eq::RWA2H} as well as the dark-state condition \eqref{eq::condDS} are in general valid for any value of $\phi$. In this paper we focus on the simplest cases $\phi=\{0,\pi\}$ where all the couplings are real. For an arbitrary phase difference, $t_{\text{CR}}^{-m}(\phi)$ and $\Upsilon_{\text{LR}}^{m,0}(\phi)$ are complex in general. Therefore, Eq. \eqref{eq::condDS} has to be fulfilled for both the real and the imaginary parts. This \textit{two} conditions make the analytical study of these phases much more involved, which is out of the purpose of this paper. We however present numerical results of the phase-difference dependence in Fig.~\ref{fig::fase} for two configurations: in Fig.~\ref{fig::fase}(a) the center dot is resonant, $\varepsilon_\text{C}-\varepsilon_\text{L,R}=3\hbar\omega$, whereas in Fig.~\ref{fig::fase}(b) it is slightly off resonance. 

As mentioned above, in resonance conditions exact cancellations of the current appear for $\phi=n\pi$, $n\in\mathbb{Z}$ if Eq.~\eqref{eq::condDS}  or Eq.~\eqref{eq:DSAC} is met for $n$ odd or even, respectively. Nevertheless, we observe in Fig.~\ref{fig::fase}(a) a substantial current drop at $\text{V}_\text{ac}/(2\hbar\omega)\approx6.3$ and $9.8$  for any $\phi$, see Fig.~\ref{fig::fase}(a). This feature is related to the zeros of the corresponding Bessel function (in this case the first and second zeros of $J_3$, $z_{1,3}$ and $z_{2,3}$).  In the off-resonant configuration the current drop at those $V_\text{ac}$ is not observed, since more sidebands are taking part in transport, cf. Fig.~\ref{fig::fase}(b). Thus the vanishing current at this conditions for $\phi=n\pi$ can only be attributed to interference effects. 
In the previous sections \ref{sec::phipi} and \ref{sec::phi0}, we have discussed that the dark states at $\phi=0$ and $\phi=\pi$ are due to different mechanisms. This becomes clear as they behave differently with a change of $\phi$. Those at $\phi=0$ are much narrower because the central symmetry depicted in Fig. \ref{Fig:esquema}(d) is broken for $\phi\neq 0$.   

Other dark states can appear for specific $\phi$ ($\ne n\pi$) which now depend in all the other parameters. This is the case of features observed around $V_\text{ac}/(2\hbar\omega)\approx2.1$ for $\phi\approx\pm 0.68\pi$ in Fig.~\ref{fig::fase}, which fulfill the dark-state condition \eqref{eq::condDS}.

For smaller amplitudes, the system is in the cotunneling regime. We find that the coherent destruction of cotunneling ---which at $\phi=\pi$ occurs for the condition $J_0(V_\text{as}/\hbar\omega)=0$, cf Sec.~\ref{sec::phipi}--- is robust and survives in a wide range of phase differences. 

%
%
%
%
\section{Conclusions}\label{sec::conc}
In summary, 
we predict quantum interferences that depend in a non-trivial way on the phase difference of the locally applied drivings. For gate voltages in phase opposition, we find destructive interferences between direct and long-range transitions which are analogous to dark states in close loop undriven triple dot molecules. 
As the edge dots levels oscillate in phase, quantum paths mediated by positive and negative detuned sidebands interfere leading to multiple dark states in the Landau-Zener-St\"uckelberg pattern.
These destructive interferences can be experimentally detected as they are of the same nature as long-range current resonances which have been unambiguously observed. 
We propose a transport configuration, where all parameters are experimentally controllable, in which these features can be measured as cancellations of the current. 
This is particularly accessible in quantum dot architectures which are within experimental reach~\cite{stehlik,braakman,floris} for both electric or magnetic field drivings. 
%
\\
\begin{acknowledgments} 
We acknowledge financial support from the Spanish Ministry MAT2014-58241-P. R.S. acknowledges support from the Spanish Ministerio de Ciencia e Innovaci\'on (MICINN) Juan de la Cierva program.
\end{acknowledgments}

\appendix
%
%


\begin{thebibliography}{99}
\bibitem{landau}
L.~D. Landau, {\it Zur Theorie der Energie\"ubertragung. II}, Phys. Z Sowjet. {\bf 2}, 46 (1932).
\bibitem{zener}
C. Zener, {\it Non-Adiabatic Crossing of Energy Levels}, \href{http://dx.doi.org/10.1098/rspa.1932.0165}{Proc. R. Soc. A {\bf 137}, 696 (1932)}.
\bibitem{stuckelberg}
E.~C.~G. St\"uckelberg, {\it Theorie der unelastischen St\"osse zwischen Atomen}, \href{https://dx.doi.org/10.5169%2Fseals-110177}{Helv. Phys. Acta {\bf 5}, 369 (1932)}.
\bibitem{majorana}
E. Majorana, {\it Atomi orientati in campo magnetico variabile}, \href{http://dx.doi.org/10.1007/BF02960953}{Nuovo Cimento {\bf 9}, 43 (1932)}.
\bibitem{oliver}
W.~D. Oliver, Y. Yu, J.~C. Lee, K.~K. Berggren, L.~S. Levitov, T.~P. Orlando, {\it Mach-Zehnder Interferometry
in a Strongly Driven Superconducting Qubit}, \href{http://www.sciencemag.org/content/310/5754/1653}{Science {\bf 310}, 1653 (2005).}
\bibitem{petta}
J.~R. Petta, H. Lu, A.~C. Gossard, {\it A Coherent Beam Splitter for Electronic Spin States}, \href{http://www.sciencemag.org/cgi/content/abstract/327/5966/669}{Science {\bf 327},669 (2010).}
\bibitem{gaudreau}
L. Gaudreau, G. Granger, A. Kam, G.~C. Aers, S.~A. Studenikin, P. Zawadzki, M. Pioro-Ladri\`ere, Z.~R. Wasilewski, A.S. Sachrajda, {\it Coherent control of three-spin states in a triple quantum dot}, \href{http://dx.doi.org/10.1038/nphys2149}{Nat. Phys. {\bf 8}, 54 (2012).}
\bibitem{quintana}
C.~M. Quintana, K.~D. Petersson, L.~W. McFaul, S.~J. Srinivasan, A.~A. Houck, and J.~R. Petta, {\it Cavity-Mediated Entanglement Generation Via Landau-Zener Interferometry}, \href{http://dx.doi.org/10.1103/PhysRevLett.110.173603}{Phys. Rev. Lett. {\bf 110}, 173603 (2013)}
\bibitem{foster}
F. Forster, G. Petersen, S. Manus, P. H\"anggi, D. Schuh, W. Wegscheider, S. Kohler, S. Ludwig, {\it Characterization of qubit dephasing by Landau-Zener-St\"uckelberg-Majorana interferometry}, \href{http://dx.doi.org/10.1103/PhysRevLett.112.116803}{Phys. Rev. Lett. 112, 116803 (2014).}
\bibitem{gloria}
G. Platero and R. Aguado, {\it Photon-assisted transport in semiconductor nanostructures}, \href{http://dx.doi.org/10.1016/j.physrep.2004.01.004}{Phys. Rep. {\bf 395}, 1 (2004).}
\bibitem{oosterkamp}
T.~H. Oosterkamp, T. Fujisawa, W.~G. van der Wiel, K. Ishibashi, R.~V. Hijman, S. Tarucha, L.~P. Kouwenhoven, {\it Microwave spectroscopy of a quantum-dot molecule}, \href{http://dx.doi.org/10.1038/27617}{Nature(London) {\bf 395}, 873 (1998).}
\bibitem{shevchenko} 
S. N. Shevchenko, S. Ashhab and F. Nori, {\it Landau-Zener-St\"uckelberg interferometry}, \href{http://dx.doi.org/10.1016/j.physrep.2010.03.002}{Phys. Reports \textbf{492}}, 1 (2010).
\bibitem{hanggi} 
F. Grossmann, T. Dittrich, P. Jung, P. Hanggi, {\it Coherent destruction of tunneling}, \href{http://dx.doi.org/10.1103/PhysRevLett.67.516}{Phys. Rev. Lett. {\bf 67}, 516 (1991).}
\bibitem{stehlik}
J. Stehlik, M.~D. Schroer, M.~Z. Maialle, M.H. Degani, J.~R. Petta, {\it Extreme Harmonic Generation in Electrically Driven Spin Resonance}, \href{http://dx.doi.org/10.1103/PhysRevLett.112.227601}{Phys. Rev. Lett. {\bf 112}, 227601 (2014).}
\bibitem{danon}
J. Danon and M.~S. Rudner, {\it Multilevel Interference Resonances in Strongly Driven Three-Level Systems}, \href{http://dx.doi.org/10.1103/PhysRevLett.113.247002}{Phys. Rev. Lett. {\bf 113} 247002 (2014).}
\bibitem{renzoni}
T. Brandes and F. Renzoni, {\it Current Switch by Coherent Trapping of Electrons in Quantum Dots}, \href{http://dx.doi.org/10.1103/PhysRevLett.85.4148}{Phys. Rev. Lett. {\bf 85}, 4148 (2000).}
\bibitem{darkbell}
R. S\'anchez and G. Platero, {\it Dark Bell states in tunnel-coupled spin qubits}, \href{http://link.aps.org/doi/10.1103/PhysRevB.87.081305}{Phys. Rev. B (Rapid Communications) 87, 081305 (2013).}
\bibitem{rogge}
M.~C. Rogge and R.J. Haug, {\it The three dimensionality of triple quantum dot stability diagrams}, \href{http://dx.doi.org/10.1088/1367-2630/11/11/113037}{New J. Phys. {\bf 11}, 113037 (2009).}
\bibitem{ghislain}
G. Granger, L. Gaudreau, A. Kam, M. Pioro-Ladri\`ere, S.A. Studenikin, Z.~R. Wasilewski, P. Zawadzki, and A.~S. Sachrajda, {\it Three-dimensional transport diagram of a triple quantum dot}, \href{http://dx.doi.org/10.1103/PhysRevB.82.075304}{Phys. Rev. B {\bf 82}, 075304 (2010).}
\bibitem{braakman}
F.~R. Braakman, P. Barthelemy, C. Reichl, W. Wegscheider, L.~M.~K. Vandersypen, {\it Photon- and phonon-assisted tunneling in the three-dimensional charge stability diagram of a triple quantum dot array}, \href{http://dx.doi.org/10.1063/1.4798335}{Appl. Phys. Lett. {\bf 102}, 112110 (2013).}
\bibitem{kiselev}
M.~N. Kiselev, K. Kikoin, M.~B. Kenmoe, {\it SU(3) Landau-Zener interferometry}, \href{http://dx.doi.org/10.1209/0295-5075/104/57004}{EPL {\bf 104}, 57004 (2013).}
\bibitem{grifoni} 
M. Grifoni, P. H\"anggi, {\it Driven quantum tunneling}, \href{http://dx.doi.org/10.1016/S0370-1573(98)00022-2}{Phys. Rep. \textbf{304}, 229 (1998).}
\bibitem{alvaro}
A. G\'omez-Le\'on and G. Platero, {\it Charge localization and dynamical spin locking in double quantum dots driven by ac magnetic fields}, \href{http://dx.doi.org/10.1103/PhysRevB.84.121310}{Phys. Rev. B {\bf 84} 121310(R) (2011).}
\bibitem{busl}
M. Busl, G. Granger, L. Gaudreau, R. S\'anchez, A. Kam, M. Pioro-Ladri\'ere, S.~A. Studenikin, P. Zawadzki, Z. R. Wasilewski,
A.~S. Sachrajda, G. Platero, {\it Bipolar spin blockade and coherent state superpositions in a triple quantum dot}, \href{http://www.nature.com/nnano/journal/vaop/ncurrent/full/nnano.2013.7.html}{Nat. Nanotechnol. 8, 261 (2013).}
\bibitem{floris}
F.~R. Braakman, P. Barthelemy, C. Reichl, W. Wegscheider, L.~M.~K. Vandersypen, {\it Long-distance coherent coupling in a quantum dot array}, \href{http://www.nature.com/nnano/journal/v8/n6/full/nnano.2013.67.html}{Nat. Nanotechnol. {\bf 8}, 432 (2013).}
\bibitem{spin}
R. S\'anchez, G. Granger, L. Gaudreau, A. Kam, M. Pioro-Ladri\`ere, S.~A. Studenikin, P. Zawadzki, A.~S. Sachrajda, G. Platero, {\it Long-range spin transfer in triple quantum dots}, \href{http://journals.aps.org/prl/abstract/10.1103/PhysRevLett.112.176803}{Phys. Rev. Lett. 112, 176803 (2014).}
\bibitem{ratner}
M. A. Ratner, {\it Bridge-assisted electron transfer: effective electronic coupling}, \href{http://dx.doi.org/10.1021/j100375a024}{J. Phys. Chem. {\bf 94}, 4877 (1990).}
\bibitem{aguado}
R. Aguado, J. I\~narrea and G. Platero, {\it Coherent resonant tunneling in ac fields}, \href{http://dx.doi.org/10.1103/PhysRevB.53.10030}{Phys. Rev. B 53, 10030 (1996).}
\bibitem{flensberg}
K. Flensberg, {\it Coherent-photon-assisted cotunneling in a Coulomb blockade device}, \href{http://dx.doi.org/10.1103/PhysRevB.55.13118}{Phys. Rev. B {\bf 55}, 13118 (1997).}
\bibitem{fernando} 
F. Gallego-Marcos, R. S{\'a}nchez and G. Platero, {\it Photon assisted long-range tunneling}, \href{http://dx.doi.org/10.1063/1.4913834}{J. Appl. Phys. \textbf{117}, 112808 (2015).}
\bibitem{stano} P. Stano, J. Klinovaja, F.R. Braakman, L. M. K. Vandersypen, D. Loss, {\it Fast Long-Distance Control of Spin Qubits by Photon Assisted Cotunneling}, \href{http://dx.doi.org/10.1103/PhysRevB.92.075302}{Phys. Rev. B {\bf 92}, 075302 (2015).}
\bibitem{michaelis}
B. Michaelis, C. Emary, C. W. J. Beenakker, {\it All-electronic coherent population trapping in quantum dots}, \href{http://dx.doi.org/10.1209/epl/i2005-10458-6}{Europhys. Lett. {\bf 73}, 677 (2006).}
\bibitem{clive} C. Emary, {\it Dark states in the magnetotransport through triple quantum dots}, \href{http://dx.doi.org/10.1103/PhysRevB.76.245319}{Phys. Rev. B \textbf{76},  245319 (2007).}
\bibitem{kobayashi}
K. Kobayashi, H. Aikawa, S. Katsumoto and Y. Iye, {\it Tuning of the Fano effect through a quantum dot in an Aharonov-Bohm interferometer}, \href{http://dx.doi.org/10.1103/PhysRevLett.88.256806}{Phys. Rev. Lett. \textbf{88}, 256806 (2002).}
\bibitem{entin}
O. Entin-Wohlman, A. Aharony, Y. Imry, and Y. Levinson, {\it The Fano effect in Aharonov-Bohm interferometers}, \href{http://dx.doi.org/10.1023/A:1013887801723}{J. Low Temp. Phys. {\bf 126}, 1251 (2002).}
\bibitem{alfredo}
A. Levy Yeyati and M. B\"uttiker, {\it Scattering phases in quantum dots: An analysis based on lattice models}, \href{http://dx.doi.org/10.1103/PhysRevB.62.7307}{Phys. Rev. B {\bf 62}, 7307 (2000).}
\bibitem{superexchange}
R. S\'anchez,  F. Gallego-Marcos, G. Platero, {\it Superexchange blockade in triple quantum dots}, \href{http://journals.aps.org/prb/abstract/10.1103/PhysRevB.89.161402}{Phys. Rev. B {\bf 89}, 161402(R) (2014).}
\bibitem{OpenQuantumSys}
H.P. Breuer and F. Petruccione, \textit{The Theory of Open Quantum Systems} (Oxford University Press, Oxford, 2002).
\bibitem{stoof}
T.~H. Stoof and Yu.~V. Nazarov,{\it Time-dependent resonant tunneling via two discrete states}, \href{http://dx.doi.org/10.1103/PhysRevB.53.1050}{Phys. Rev. B {\bf 53}, 1050 (1996).}
\end{thebibliography}
\end{document}